\documentclass[journal]{IEEEtran}
\ifCLASSINFOpdf
\else
\fi
\usepackage{amsfonts}
\usepackage{amsmath}
\usepackage{amssymb}
\usepackage{graphicx}
\usepackage{algorithm}
\usepackage{algpseudocode}
\usepackage{array}
\usepackage{booktabs}
\usepackage{makecell}
\usepackage{cite}

\ifCLASSOPTIONcompsoc
\usepackage[caption=false,font=normalsize,labelfon
t=sf,textfont=sf]{subfig}
\else
\usepackage[caption=false,font=footnotesize]{subfi
g}
\fi

\begin{document}
%
\title{Radial Coverage Strength for Optimization of Multi-Camera Deployment}
%
%
%

\author{Zike~Lei,
        Xi~Chen*,~\IEEEmembership{Member,~IEEE,}
        Xiang~Chen,~\IEEEmembership{Member,~IEEE,}
        Li~Chai,~\IEEEmembership{Member,~IEEE}
\thanks{*This work was supported under the National Natural Science Foundation
of China under Grant 61703315, 61625305.}%
\thanks{Zike Lei is with the School of Information Science and Engineering, Wuhan University of Science and Technology, Hubei, 430081 China (e-mail: leizike@wust.edu.cn).}
\thanks{Xi Chen is with the Engineering Research Center of Metallurgical Automation and Measurement Technology, Wuhan University of Science and Technology, Hubei 430081, China (e-mail: chenxi\_99@wust.edu.cn) (corresponding author).}
\thanks{Xiang Chen is with the Department of Electrical and Computer Engineering, University of Windsor, Ontario, N9B 3P4 Canada (e-mail: xchen@uwindsor.ca).}
\thanks{Li Chai is with the Engineering Research Center of Metallurgical Automation and Measurement Technology, Wuhan University of Science and Technology, Hubei, 430081 China (e-mail: chaili@wust.edu.cn).}
}
\maketitle

\begin{abstract}

In this paper, a new concept, radial coverage strength, is first proposed to characterize the visual sensing performance when the orientation of the target pose is considered. In particular, the elevation angle of the optical pose of the visual sensor is taken to decompose the visual coverage strength into effective and ineffective components, motivated by the imaging intuition. An optimization problem is then formulated for a multi-camera network to maximize the coverage of the object area based on the strength information fusion along the effective coverage strength direction through the deployment of the angle between radial coverage vector of the camera optical pose. Both simulation and experiments are conducted to validate the proposed approach and comparison with existing methods is also provided.  


\end{abstract}

\begin{IEEEkeywords}
Multi-camera network, data fusion, radial coverage, visual sensor deployment.
\end{IEEEkeywords}

%
\IEEEpeerreviewmaketitle

\section{Introduction}
\label{section1}
%
%
%
%
\IEEEPARstart{C}{ameras}, as a typical non-contact field sensor, are widely used in occasions that require interaction with ambient target objectives, such as unmanned system formation \cite{li2018compatible, Aranda2015formation}, visual servoing and tracking \cite{stavnitzky2000multiple, zhang2011motion,wang2011new}, robotic localization and navigation \cite{Wang2014A, Panahandeh2014Vision}, etc. While the camera sensor is welcomed due to many advantages such as compact size, low power consumption, reasonably low cost, and rich information, the limitation of a camera sensor is also obvious, for example, in terms of its limited field of view and occlusion. In many practical applications such as visual coverage \cite{zhang20153, zhang2018visual, jesus2018computing}, objective reconstruction \cite{chen2005vision,mavrinac2015semiautomatic}, and large-scale surveillance \cite{mavrinac2014coverage}, hence, multi-camera networks are highly desired to allow individual cameras to perform collaboratively, mostly, through spatial deployment. Therefore, how to optimize the spatial deployment of camera sensor networks becomes a very interesting yet important problem that attracts a lot of attentions in recent years \cite{Guvensan2011survey}. In this regard, various coverage models have been proposed for visual field sensors, such as 2-D circular sector sensing in \cite{gusrialdi2008coverage, zhang2016coverage, parapari2016distributed, zhang2015distributed} and fish-eye cameras in \cite{li2018compatible}, 3-D coverage models taking into account the intrinsic and/or extrinsic parameters of cameras, such as resolution, field of view (FOV), focus and occlusion, etc., in \cite{Jiang2010A, mavrinac2014coverage, zhang2018visual, zhang20153}. In particular, a comprehensive 3-D coverage model with continuous measure is introduced in \cite{mavrinac2014coverage} and is further extended and successfully applied in many scenarios in \cite{zhang20153} which considers both depth distance and view angle for a better physical interpretation. Besides, a `visual distance' is also proposed in \cite{zhang2018visual} to characterize the pose difference between a single camera and a target point, and the optimal deployment of multiple cameras is solved based on this performance measure accordingly. The advantage of the coverage model in \cite{mavrinac2014coverage,zhang20153} lies in the fact that it presents an effective 3-D parameterized region within which the visual sensing quality can be continuously measured. However, the orientations of both camera and target object poses are not considered in this model, instead, they are actually the main feature considered in the `visual distance' model \cite{zhang2018visual}. There are also some other work based on data fusion which combine the coverage strength of multiple 2-D camera sensors via some mathematic operations \cite{xu2018novel, Shan2016Target} but the deployment is considered. 
Fairly speaking, the image fusion techniques are very mature nowadays \cite{wang2005comparative} but, again, there lacks research on the application of these techniques to camera sensor deployment.  

On the other hand, saving costs of implementation is always the most important goal of industrial applications. In various field applications with camera sensor networks, maximizing the coverage area is a very interesting and important task which would lead to const-effective solutions to camera sensor network deployment. In this paper, a new concept, `radial coverage strength', is proposed and facilitated with the advantages of both the coverage strength model and the visual distance model developed in \cite{mavrinac2014coverage,zhang20153,zhang2018visual}. This new concept allows one to characterize a parameterized coverage region while the relative pose between the camera sensor and the target area is also considered. An optimization problem is then formulated by applying the radial coverage strength as the performance measure through information fusion to tackle the task of maximizing the coverage area for the camera sensor network. Both simulation and experiments are presented to validate the main results, as well as comparison with some existing methods. 
 
The organization of the remaining parts of this paper is as follows: some background information is presented for the object model and camera model in Section \ref{section2} as preliminaries; the radial coverage strength is proposed and described in Section \ref{section3}; in Section \ref{section4}, the fusing mechanism for the radial coverage strength is presented which lays down the foundation for the deployment optimization of the camera sensor network featured in Section \ref{section5}; simulation and experiments are provided in Section \ref{section6} to validate the effectiveness of the proposed approach; Section \ref{section7} concludes the paper.

\emph{Notations:} $\mathbb{R}^n$ denotes $n$-dimensional Euclidean space, $\mathbb{R}^+$ denotes positive real number, and $\mathbb{N}^+$ represents a positive integer. $SO(3)$ denotes the Special Orthogonal Group in  the 3-D space. $\|\cdot\|$ stands for the Euclidean norm. Let $\boldsymbol{\upsilon}_1,\boldsymbol{\upsilon}_2\in\mathbb{R}^n$ be $n$-dimensional vectors, then $\mathrm{Proj}(\boldsymbol{\upsilon}_1,\boldsymbol{\upsilon}_2) = \boldsymbol{\upsilon}_2 - (\boldsymbol{\upsilon}_1^\mathrm{T}\boldsymbol{\upsilon}_2/\|\boldsymbol{\upsilon}_1\|^2)\boldsymbol{\upsilon}_1$ denote the projection of $\upsilon_2$ on a plane with $\upsilon_1$ being its normal vector.

\section{Preliminaries}
\label{section2}
In this section, some preliminary knowledge such as frame transformation, environment model, camera model and relevance model are introduced respectively to describe the camera network as well as the 3-D environment to be covered.

\subsection{World Frame and Local Frame}
\begin{figure}[htbp]
\centering
\includegraphics[scale=1]{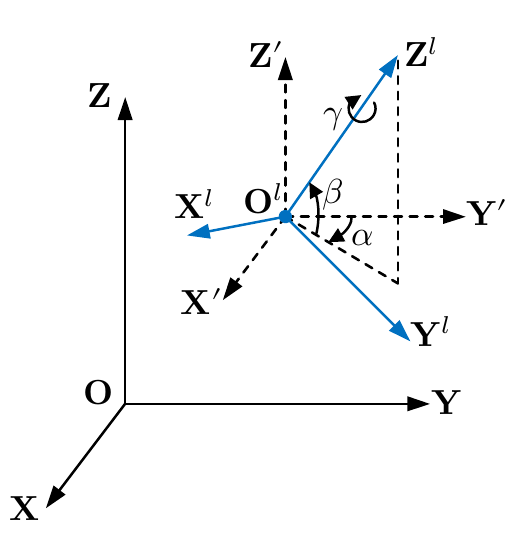}
\caption{World frame and local frame.}
\label{S2_frame}
\end{figure}

As shown in Fig. \ref{S2_frame}, the world frame $\mathcal{F}^w$ is represented by axis $\mathbf{X}\mathbf{Y}\mathbf{Z}$ with $\mathbf{O}$ being its origin, the local frame $\mathcal{F}^l$ is represented by axis $\mathbf{X}^l\mathbf{Y}^l\mathbf{Z}^l$ with $\mathbf{O}^l$ being its origin. The axis $\mathbf{X}'\mathbf{Y}'\mathbf{Z}'$ indicated by the dotted line is the reference axis parallel to $\mathbf{X}\mathbf{Y}\mathbf{Z}$. 

Let $\mathbf{c}\in\mathbb{R}^6$ denote the coordinates of $\mathcal{F}^l$:
\begin{equation}
\label{c}
\mathbf{c} = [\begin{IEEEeqnarraybox*}[][c]{,c/c,}
\boldsymbol{\varsigma}^\mathrm{T} & \boldsymbol{\vartheta}^\mathrm{T}
\end{IEEEeqnarraybox*}]^\mathrm{T}
= [\begin{IEEEeqnarraybox*}[][c]{,c/c/c/c/c/c,}
x_l & y_l & z_l & \alpha & \beta & \gamma
\end{IEEEeqnarraybox*}]^\mathrm{T}
\end{equation}
where the position component
$
\boldsymbol{\varsigma} = [\begin{IEEEeqnarraybox*}[][c]{,c/c/c,}
x_l & y_l & z_l
\end{IEEEeqnarraybox*}]^\mathrm{T}\in\mathbb{R}^3
$
denotes the coordinates of $\mathbf{O}^l$ in $\mathcal{F}^w$, and the orientation component
$
\boldsymbol{\vartheta} = [\begin{IEEEeqnarraybox*}[][c]{,c/c/c,}
\alpha & \beta & \gamma
\end{IEEEeqnarraybox*}]^\mathrm{T}\in\mathbb{R}^3
$
denote the orientation of $\mathcal{F}^l$, where $\alpha\in[-\pi,\pi)$ is the yaw angle, $\beta\in[-\pi/2,\pi/2]$ is the pitch angle, and $\gamma\in[-\pi/2,\pi/2]$ is the roll angle of $\mathcal{F}^l$ measured in $\mathcal{F}^w$. For a point
$
\mathbf{s} = [\begin{IEEEeqnarraybox*}[][c]{,c/c/c,}
x & y & z
\end{IEEEeqnarraybox*}]^\mathrm{T}\in\mathbb{R}^3
$ in $\mathcal{F}^w$, let its coordinate in $\mathcal{F}^l$ be
$
\mathbf{s}^l = [\begin{IEEEeqnarraybox*}[][c]{,c/c/c,}
x^l & y^l & z^l
\end{IEEEeqnarraybox*}]^\mathrm{T}\in\mathbb{R}^3
$
, then the following relationship holds \cite{zhang2000flexible}:
\begin{equation}
\label{s^l}
\mathbf{s}^l = \mathbf{R}(\mathbf{s}-\boldsymbol{\varsigma})
\end{equation}
where
\begin{equation}
\label{R}
\mathbf{R} = \mathbf{R}^{\gamma}\mathbf{R}^{\beta}\mathbf{R}^{\alpha}\left[\begin{IEEEeqnarraybox*}[][c]{,c/c/c,}
1 & 0 & 0 \\
0 & 0 & -1 \\
0 & 1 & 0
\end{IEEEeqnarraybox*}\right]\in SO(3)
\end{equation}
is the rotation transformation matrix from $\mathcal{F}^w$ to $\mathcal{F}^l$, and
\begin{align}
\label{R^alpha}
\mathbf{R}^{\alpha} &= \left[\begin{IEEEeqnarraybox*}[][c]{,c/c/c,}
\cos{\alpha} & 0 & \sin{\alpha} \\
0 & 1 & 0 \\
-\sin{\alpha} & 0 & \cos{\alpha}
\end{IEEEeqnarraybox*}\right]\in SO(3) \\
\label{R^beta}
\mathbf{R}^{\beta} &= \left[\begin{IEEEeqnarraybox*}[][c]{,c/c/c,}
1 & 0 & 0 \\
0 & \cos{\beta} & -\sin{\beta} \\
0 & \sin{\beta} & \cos{\beta}
\end{IEEEeqnarraybox*}\right]\in SO(3) \\
\label{R^gamma}
\mathbf{R}^{\gamma} &= \left[\begin{IEEEeqnarraybox*}[][c]{,c/c/c,}
\cos{\gamma} & -\sin{\gamma} & 0 \\
\sin{\gamma} & \cos{\gamma} & 0 \\
0 & 0 & 1
\end{IEEEeqnarraybox*}\right]\in SO(3)
\end{align}
are the rotation transformation matrices from $\mathcal{F}^w$ to $\mathcal{F}^l$ for yaw, pitch and roll, respectively.

\subsection{Object Model}
The object model is introduced in $\mathcal{F}^w$. The object model represents the area of the object to be covered by cameras and is characterized with meshed triangle pieces. Each triangle piece is regarded to be an atomic unit in analysis. The area of each triangle piece is set to be smaller than a given threshold $\sigma$. Moreover, let $\tau_k$ denote the $k^{th}$ triangle piece ($k=1,2,\cdots, K$), with $K\in\mathbb{N}^+$ being the total number of the triangle pieces in the 3-D object model.

For each triangle piece, a local frame is defined on it. The origin of the frame is defined as the center of the triangle piece and $\mathbf{Z}^l$-axis is along the front face normal direction of the triangle piece. Let the directional point $\mathbf{p}_k\in\mathbb{R}^6$ denote the coordinates of the unit normal vector of $\tau_k$:
\begin{equation}
\label{p_k}
\mathbf{p}_k = [\begin{IEEEeqnarraybox*}[][c]{,c/c,}
\mathbf{s}^\mathrm{T}_k & \boldsymbol{\theta}^\mathrm{T}_k
\end{IEEEeqnarraybox*}]^\mathrm{T}
= [\begin{IEEEeqnarraybox*}[][c]{,c/c/c/c/c/c,}
x_k & y_k & z_k & \rho_k & \eta_k & \mu_k
\end{IEEEeqnarraybox*}]^\mathrm{T}
\end{equation}
where
$
\mathbf{s}_k = [\begin{IEEEeqnarraybox*}[][c]{,c/c/c,}
x_k & y_k & z_k
\end{IEEEeqnarraybox*}]^\mathrm{T}\in\mathbb{R}^3
$
is the position component and
$
\boldsymbol{\theta}_k = [\begin{IEEEeqnarraybox*}[][c]{,c/c/c,}
\rho_k & \eta_k & \mu_k
\end{IEEEeqnarraybox*}]^\mathrm{T}\in\mathbb{R}^3
$
is the orientation component. Then the front face normal direction $\mathbf{n}_k$ of $\tau_k$ can be characterized by:
\begin{equation}
\label{n_k}
\mathbf{n}_k = [\begin{IEEEeqnarraybox*}[][c]{,c/c/c,}
\sin{\rho_k}\cos{\eta_k} & \;\sin{\rho_k}\sin{\eta_k} & \;\cos{\rho_k}
\end{IEEEeqnarraybox*}]^\mathrm{T}
\end{equation}
in $\mathcal{F}^w$.

In this paper, the directional point is used to represents the triangle piece. Let $\Omega=\{\mathbf{p}_1,\mathbf{p}_2,\cdots,\mathbf{p}_K\}$ be the set of all the directional points of triangle piece. For the specified problem in this paper, it is assumed that the triangle-piece model of the object is available for the coverage, for example, the CAD file of a part. Besides, for each triangle piece, a relevance weight is assigned to indicate its importance in the coverage. In this work, the relevance weight is taken as the area of the triangle piece:
\begin{equation}
\label{Rel}
Rel(\mathbf{p}_k) = Area(\tau_k)
\end{equation}

\subsection{Camera Model}
In this paper, the pinhole camera is used to perform the coverage task. The camera model consists of both intrinsic parameters and extrinsic parameters \cite{zhang2000flexible}. Intrinsic parameters are shown in Table \ref{t1} which are determined by the camera itself.

Extrinsic parameters include the position and the orientation component. For each camera, a local frame is defined on it, called the camera frame $\mathcal{F}^c$. The origin of the frame is defined at the optical center of the camera and $\mathbf{Z}^c$-axis is along the optical axis of the camera. The $\mathbf{X}^c$-axis and $\mathbf{Y}^c$-axis are along the opposite directions of $\mathbf{U}^c$-axis and $\mathbf{V}^c$-axis axes of the Charge Coupled Device (CCD), respectively, which form a right-handed frame together with $\mathbf{Z}^c$-axis. $\mathcal{F}^{ci}$ is used to represent the camera frame of the $i^{th}$ camera. Let $\mathbf{c}_i\in\mathbb{R}^6$ denote the coordinates of the position of the $i^{th}$ camera in $\mathcal{F}^w$ ($i=1,2,\cdots, N$), which is the extrinsic parameters, with $N\in\mathbb{N}^+$ being the total number of the cameras in the task:
\begin{equation}
\label{c_i}
\mathbf{c}_i = [\begin{IEEEeqnarraybox*}[][c]{,c/c,}
\boldsymbol{\varsigma}^\mathrm{T}_i & \boldsymbol{\vartheta}^\mathrm{T}_i
\end{IEEEeqnarraybox*}]^\mathrm{T}
= [\begin{IEEEeqnarraybox*}[][c]{,c/c/c/c/c/c,}
x_i & y_i & z_i & \alpha_i & \beta_i & \gamma_i
\end{IEEEeqnarraybox*}]^\mathrm{T}
\end{equation}
where
$
\boldsymbol{\varsigma}_i = [\begin{IEEEeqnarraybox*}[][c]{,c/c/c,}
x_i & y_i & z_i
\end{IEEEeqnarraybox*}]^\mathrm{T}\in\mathbb{R}^3
$
is the position component of the $i^{th}$ camera and
$
\boldsymbol{\vartheta}_i = [\begin{IEEEeqnarraybox*}[][c]{,c/c/c,}
\alpha_i & \beta_i & \gamma_i
\end{IEEEeqnarraybox*}]^\mathrm{T}\in\mathbb{R}^3
$
is the orientation component of the $i^{th}$ camera. Let $\Gamma=\{\mathbf{c}_1,\mathbf{c}_2,\cdots,\mathbf{c}_N\}$ be the set of extrinsic parameters for all cameras.

It is noted that, from equation (\ref{c}), the 3-D coordinate $\mathbf{s}_k$ of $\tau_k$ in $\mathcal{F}^w$ can be transformed into the coordinate
$
\mathbf{s}^{ci}_k = [\begin{IEEEeqnarraybox*}[][c]{,c/c/c,}
x^{ci}_k & y^{ci}_k & z^{ci}_k
\end{IEEEeqnarraybox*}]^\mathrm{T}\in\mathbb{R}^3
$
in $\mathcal{F}^{ci}$:
\begin{equation}
\label{s^ci_k}
\mathbf{s}^{ci}_k = \mathbf{R}_i(\mathbf{s}_k-\boldsymbol{\varsigma}_i)
\end{equation}
where $\mathbf{R}_i\in SO(3)$ is the rotation transformation matrix from $\mathcal{F}^w$ to $\mathcal{F}^{ci}$.

\begin{table}[!t]\label{intrinsic parameter}
\renewcommand{\arraystretch}{1.3}
\caption{Intrinsic Parameters of Camera Model}
\label{t1}
\centering
\begin{tabular}{p{62pt}l}
\hline\hline
\specialrule{0em}{0pt}{2pt}
Parameter & Description \\
\specialrule{0em}{2pt}{0pt}
\hline
\specialrule{0em}{2pt}{0pt}
$f\in\mathbb{R}^+$ & Lens focal length (mm) \\
$s_u\in\mathbb{R}^+$ & Horizontal pixel dimensions (mm/pixel) \\
$s_v\in\mathbb{R}^+$ & Vertical pixel dimensions (mm/pixel) \\
$
\mathbf{o} \!=\! [\begin{IEEEeqnarraybox*}[][c]{,c/c,}
\!o_u\! & o_v\!
\end{IEEEeqnarraybox*}]^\mathrm{T}\!\!\in\!\mathbb{R}^2
$
& Principle point (pixel) \\
$w\in\mathbb{R}^+$ & Image width (pixel) \\
$h\in\mathbb{R}^+$ & Image height (pixel) \\
$d_a\in\mathbb{R}^+$ & Effective aperture diameter of optical lens (mm) \\
$d_s\in\mathbb{R}^+$ & Focusing distance (mm) \\
$\varphi_t,\varphi_b\in[0,\pi/2)$ & FOV angles to top / bottom image boundary (rad) \\
$\varphi_l,\varphi_r\in[0,\pi/2)$ & FOV angles to left / right image boundary (rad) \\
\specialrule{0em}{2pt}{0pt}
\hline\hline
\end{tabular}
\end{table}

\section{Radial Coverage Strength}
\label{section3}
In this section, a new criteria called the Radial Coverage Strength is proposed to characterize the coverage performance of single triangle piece in single camera, involving the resolution, FOV, focus, and occlusion, taking $\mathbf{s}^{ci}_k$, $\mathbf{n}_k$ and the intrinsic parameters of the camera into account.

\subsection{Resolution}
Resolution is an important criterion of the coverage performance of triangle pieces in the camera, which describes the detail an image holds for the target object. The resolution criterion of $\mathbf{p}_k$ under the $i^{th}$ camera is defined as
\begin{equation}
\label{C^(R)_i}
C^{R}_i(\mathbf{p}_k) = \frac{fd_s}{(d_s-f)z^{ci}_k\max(s_u,s_v)}
\end{equation}
where $f$, $d_s$, $s_u$ and $s_v$ are intrinsic parameters of the camera that explained in Table I, $z_k^{ci}$ is the coordinate of $\mathbf{p}_k$ on $\mathbf{Z}^{ci}$-axis.

\subsection{Field of View}
\begin{figure}[!t]
\centering
\includegraphics[scale=1]{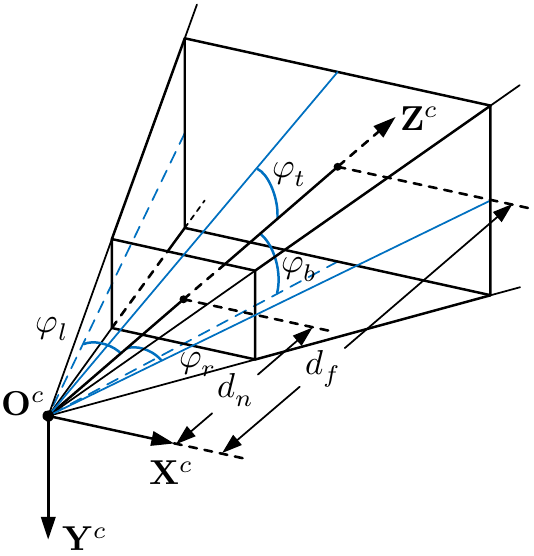}
\caption{Illustration for FOV and focus parameters.}
\label{S2_fov_focus}
\end{figure}

As shown in Fig. \ref{S2_fov_focus}, the blue rays are the angle bisectors of the left, right, upper, and lower boundaries of the field of view, $\varphi_l$, $\varphi_r$ are the horizontal FOV angles, and $\varphi_t$, $\varphi_b$ are the vertical FOV angles. These angles can be determined by the camera intrinsic parameters.

For the directional point $\mathbf{p}_k$, if the following geometrical constraints are satisfied
\begin{equation}
\label{fov1}
z^{ci}_k \ge 0
\end{equation}
\begin{equation}
\label{fov2}
-\tan{\varphi_l} \le \frac{x^{ci}_k}{z^{ci}_k} \le \tan{\varphi_r}
\end{equation}
\begin{equation}
\label{fov3}
-\tan{\varphi_t} \le \frac{y^{ci}_k}{z^{ci}_k} \le \tan{\varphi_b}
\end{equation}
then it is said that $\mathbf{p}_k$ falls in the FOV of the $i^{th}$ camera.

The FOV criterion of $\mathbf{p}_k$ is defined as
\begin{equation}
\label{C^FOV_i}
C^{FOV}_i(\mathbf{p}_k) =
\begin{cases}
1 & \text{fall in FOV} \\
0 & \text{otherwise}
\end{cases}
\end{equation}

\subsection{Focus}
The imaging of the object is blurred if it is not on the focal plane of the camera. However, the degree of defocus is permitted within a certain range. As shown in Fig. \ref{S2_fov_focus}, four boundaries of the FOV form a pyramid, and it is divided into a frustum by a near-field depth plane and a far-field depth plane, which is called the view frustum.

For the directional point $\mathbf{p}_k$, given diameter $\delta$ of the permissible circle of confusion in pixel, if $z_k^{ci}$ falls in the following range
\begin{equation}
\label{focus1}
d_n \le z^{ci}_k \le d_f
\end{equation}
then it is said that $\mathbf{p}_k$ is focused by the $i^{th}$ camera, where
\begin{equation}
\label{d_n}
d_n = \frac{d_a d_s f}{d_a f+\delta \min(s_u,s_v)(d_s-f)}
\end{equation}
\begin{equation}
\label{d_f}
d_f = \frac{d_a d_s f}{d_a f-\delta \min(s_u,s_v)(d_s-f)}
\end{equation}
are the near-field depth and far-field depth, respectively, with $d_n < d_f$.

The focus criterion of $\mathbf{p}_k$ is defined as
\begin{equation}
\label{C^F_i}
C^{F}_i(\mathbf{p}_k) =
\begin{cases}
1 & \text{focused} \\
0 & \text{otherwise}
\end{cases}
\end{equation}

\subsection{Occlusion}
In this paper, internal occlusion and external occlusion are both considered. Internal occlusion is from the object model itself, that is some other triangle pieces exist on the segment connecting $\mathbf{p}_k$ and the optical center $\mathbf{O}^{ci}$ of the $i^{th}$ camera. External occlusion is that the obstacles existing around the object model, which also affects the deployment of the cameras. In Section VI, a case with external occlusion is shown by simulation to illustrate the occlusion handling.

\begin{figure}[!t]
\centering
\includegraphics[scale=1]{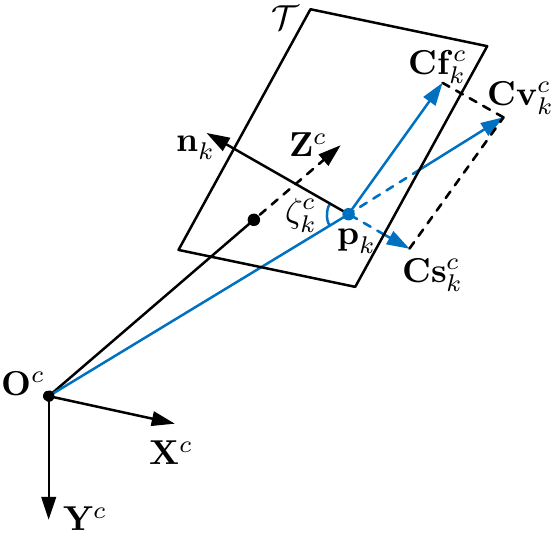}
\caption{Illustration for radial coverage vector.}
\label{S3_radial_coverage_strength}
\end{figure}

As shown in Fig. \ref{S3_radial_coverage_strength}, the angle between vector $\mathbf{n}_k$ and vector $\mathbf{p}_k\mathbf{O}^{ci}$ is defined as $\zeta^{ci}_k\in[0,\pi]$, called the elevation angle between $\tau_k$ and the $i^{th}$ camera, then
\begin{equation}
\label{zeta^ci_k}
\zeta^{ci}_k = \arccos{\frac{\mathbf{n}_k^\mathrm{T}(\boldsymbol{\varsigma}_i-\mathbf{s}_k)}{\|\boldsymbol{\varsigma}_i-\mathbf{s}_k\|}}
\end{equation}
The directional point $\mathbf{p}_k$ is considered occluded in the $i^{th}$ camera if any of the following conditions are met
\begin{enumerate}[\IEEEsetlabelwidth{12)}]
\item
$\zeta^{ci}_k\ge\pi/2$, means the camera cannot see the front face of the triangle piece.
\item
The segments between the camera optical center $\mathbf{O}^{ci}$ and the three vertices of $\tau_k$ passes through other triangle pieces or obstacles.
\end{enumerate}

Then the occlusion criterion of $\mathbf{p}_k$ is defined as
\begin{equation}
\label{C^O_i}
C^{O}_i(\mathbf{p}_k) =
\begin{cases}
0 & \text{occluded} \\
1 & \text{otherwise}
\end{cases}
\end{equation}

\subsection{Radial Coverage Vector}
\label{subsection3e}
The radial coverage vector is proposed to represent the coverage performance of a single triangle piece in a single camera. Let $\mathbf{Cv}^{ci}_k\in\mathbb{R}^3$ denote the radial coverage vector of $\mathbf{p}_k$ under the $i^{th}$ camera in $\mathcal{F}^w$:
\begin{equation}
\label{CV^ci_k}
\mathbf{Cv}^{ci}_k = \frac{C^{R}_i(\mathbf{p}_k)C^{FOV}_i(\mathbf{p}_k)C^{F}_i(\mathbf{p}_k)C^{O}_i(\mathbf{p}_k)(\mathbf{s}_k-\boldsymbol{\varsigma}_i)}{\|\mathbf{s}_k-\boldsymbol{\varsigma}_i\|}
\end{equation}
As shown in Fig. \ref{S3_radial_coverage_strength}, the light is illuminated from $\mathbf{p}_k$ to camera optical center $\mathbf{O}^c$, then we can assume that the camera ``sees'' the triangle piece along the opposite direction of the light, that is, the $\mathbf{Cv}^c_k$ direction. The length of $\mathbf{Cv}^c_k$ takes into account multiple visual criteria. The radial coverage vector can be decomposed into a vector $\mathbf{Cf}^c_k$ parallel to the triangle piece plane and a vector $\mathbf{Cs}^c_k$ parallel to the front face normal direction of the triangle piece. Let $\mathbf{Cf}^{ci}_k\in\mathbb{R}^3$ denote the fusion vector of $\mathbf{p}_k$ under the $i^{th}$ camera in $\mathcal{F}^w$:
\begin{equation}
\label{CF^ci_k}
\mathbf{Cf}^{ci}_k = \mathrm{Proj}(\mathbf{n}_k,\mathbf{Cv}^{ci}_k)
\end{equation}
which provides the direction information required for further fusion algorithm. Let $\mathbf{Cs}^{ci}_k\in\mathbb{R}^3$ denote the effective radial coverage vector of $\mathbf{p}_k$ under the $i^{th}$ camera in $\mathcal{F}^w$:
\begin{equation}
\label{CS^ci_k}
\mathbf{Cs}^{ci}_k = \mathbf{Cv}^{ci}_k - \mathbf{Cf}^{ci}_k
\end{equation}
Since the effective radial coverage vector is parallel to the normal direction of the triangle piece, it represents a component of the camera that see the triangle piece vertically, providing the coverage strength information. Therefore, the effective radial coverage strength of $\mathbf{p}_k$ under the $i^{th}$ camera is defined as:
\begin{equation}
\label{abs_CS^ci_k}
\|\mathbf{Cs}^{ci}_k\| = \cos{\zeta^{ci}_k}\|\mathbf{Cv}^{ci}_k\|
\end{equation}
where $\cos{\zeta^{ci}_k}$ represents the reduction of the coverage, since the orientation of the triangle pieces must be considered.

\begin{table}[!t]
\renewcommand{\arraystretch}{1.3}
\caption{Parameters of Improved Genetic Algorithm}
\label{t2}
\centering
\begin{tabular}{ll}
\hline\hline
\specialrule{0em}{0pt}{2pt}
Parameter & Description \\
\specialrule{0em}{2pt}{0pt}
\hline
\specialrule{0em}{2pt}{0pt}
$M\in\mathbb{N}^+$ & Size of population \\
$L\in\mathbb{N}^+$ & Length of chromosome (number of genes) \\
$\Upsilon_{min},\Upsilon_{max}\in\mathbb{N}^+$ & Minimum / Maximum length of recombination \\
$\Psi\in(0,1]$ & Mutation probability of each gene \\
$T$ & Algorithm termination condition \\
\specialrule{0em}{2pt}{0pt}
\hline\hline
\end{tabular}
\end{table}

\section{Fused Coverage Strength}
\label{section4}
In the existing literatures, some coverage strength algorithms have been proposed to model the coverage performance by cameras in \cite{gusrialdi2008coverage,zhang2015distributed,zhang2018visual,zhang20153,mavrinac2015semiautomatic}. The coverage strength algorithms in \cite{gusrialdi2008coverage}, \cite{zhang2015distributed} and \cite{zhang2018visual} are constructed from mathematic perspective, thus be lack of enough factual basis. To reach the goals in industry, some coverage strength algorithms are proposed from physic perspective for combining the visual criteria for resolution, FOV, focus and occlusion in \cite{zhang20153} and \cite{mavrinac2015semiautomatic}. However, those coverage strength only consider the single camera, but never the data fusion of multiple cameras for cooperations. In this section, an effective fused coverage strength algorithm is proposed to enhance the overall coverage performance under the multi-camera network.

The effective radial coverage strength and the fusion vector reveals the detailed performance of the triangle piece captured by single camera. For $\tau_k$, since all fusion vectors are coplanar, the fusion algorithm are always calculated on a 2-D plane, In this respect, only two cameras are selected for each calculation of fusion, representing the coverage performance in two directions. On the basis of this idea, a fused coverage strength matrix of $\mathbf{p}_k$ is constructed as
\begin{equation}
\label{Lambda_k}
\mathbf{\Lambda}_k = \left[\begin{IEEEeqnarraybox*}[][c]{,c/c/c/c,}
Cs_k(c_1,c_1) & Cs_k(c_1,c_2) & \cdots & Cs_k(c_1,c_N) \\
Cs_k(c_2,c_1) & Cs_k(c_2,c_2) & \cdots & Cs_k(c_2,c_N) \\
\vdots & \vdots & \ddots & \vdots \\
Cs_k(c_N,c_1) & Cs_k(c_N,c_2) & \cdots & Cs_k(c_N,c_N)
\end{IEEEeqnarraybox*}\right]
\end{equation}
where
\begin{equation}
\label{Cs^cicj_k}
\begin{aligned}
&Cs_k(c_i,c_j) = \\
&\begin{cases}
\max(\|\mathbf{Cs}^{ci}_k\|, \|\mathbf{Cs}^{cj}_k\|) \qquad \;\;\;\;\;\;\;\;\;\;\;\,\,\, \mathbf{Cf}^{ci}_k=\mathbf{0}\text{ or }\mathbf{Cf}^{cj}_k=\mathbf{0} \\[3pt]
\displaystyle
\left\|\frac{\|\mathbf{Cs}^{ci}_k\|\mathbf{Cf}^{ci}_k}{\|\mathbf{Cf}^{ci}_k\|}
\!+\!
\mathrm{Proj}\!\left(\mathbf{Cf}^{ci}_k,\frac{\|\mathbf{Cs}^{cj}_k\|\mathbf{Cf}^{cj}_k}{\|\mathbf{Cf}^{cj}_k\|}\right)\right\| \;\;\;\,\,\,\, \text{otherwise}
\end{cases}
\end{aligned}
\end{equation}
denotes the fused coverage strength of $\mathbf{p}_k$ under the $i^{th}$ camera and the $j^{th}$ camera for $i,j=1,2,\cdots, N$. Then the fused coverage strength of $\mathbf{p}_k$ under the multi-camera network is selected as the maximum element in $\mathbf{\Lambda}_k$:
\begin{equation}
\label{Cs_p_k}
Cs(\mathbf{p}_k) = \mathop{\max}_{i,j}(Cs_k(c_i,c_j))
\end{equation}

\section{Cameras Deployment Optimization}
\label{section5}
The goal of multi-camera deployment is to find a set of appropriate extrinsic parameters of the cameras to optimize the tasks. In the existing literatures, different optimization algorithms have been applied for various tasks, such as gradient descent algorithm (GD), particle swarm optimization algorithm (PSO), standard genetic algorithm (SGA), ant colony optimization algorithm (ACO) and simulated annealing algorithm (SA). In this paper, an improved genetic algorithm (IGA) is proposed to optimize the camera deployment problem.

\subsection{Task Formulation}
Before construct the cost function, a definition is given as follows to state whether a triangle piece is recognized by the camera network. Given a coverage strength threshold $thold\in\mathbb{R}^+$, if the fused coverage strength of $\mathbf{p}_k$ satisfies
\begin{equation}
\label{thold}
Cs(\mathbf{p}_k) \ge thold
\end{equation}
then $\mathbf{p}_k$ is said to be recognized by the multi-camera network.

The cost function can be formulated as
\begin{equation}
\label{H}
\begin{aligned}
\mathcal{H}(\mathbf{c}_1,\cdots&,\mathbf{c}_N,\mathbf{p}_1,\cdots,\mathbf{p}_K) = \\
&\sum^K_{k=1}Reco(\mathbf{c}_1,\cdots,\mathbf{c}_N,\mathbf{p}_k)Rel(\mathbf{p}_k)
\end{aligned}
\end{equation}
which represents the total area of the recognized triangle pieces, where
\begin{equation}
\label{Reco}
Reco(\mathbf{c}_1,\cdots,\mathbf{c}_N,\mathbf{p}_k) =
\begin{cases}
1 & \text{recognized} \\
0 & \text{otherwise}
\end{cases}
\end{equation}
is an criterion for recognition. Hence, the coverage optimization problem of the camera network can be formulated as
\begin{equation}
\label{argmaxH}
\begin{aligned}
\mathop{\arg\max}_{\mathbf{c}_1,\cdots,\mathbf{c}_N}&~\mathcal{H}(\mathbf{c}_1,\cdots,\mathbf{c}_N,\mathbf{p}_1,\cdots,\mathbf{p}_K)\\
&s.t. ~~\mathbf{c}_i\in\Gamma,\mathbf{p}_k\in\Omega
\end{aligned}
\end{equation}

\begin{figure}
\begin{algorithm}[H]
\caption{Improved Genetic Algorithm for Optimization}
\label{algorithm1}
\textbf{Input:} $\Omega$, $IGA$, $f$, $s_u$, $s_v$, $d_a$, $d_s$, $\varphi_l$, $\varphi_r$, $\varphi_t$, $\varphi_b$, $\delta$\\
\textbf{Output:} $\mathbf{c}_1$, $\mathbf{c}_2$, $\cdots$, $\mathbf{c}_N$
\makeatletter
\newcommand{\algmargin}{\the\ALG@thistlm}
\makeatother
\algnewcommand{\parState}[1]{\State%
  \parbox[t]{\dimexpr\linewidth-\algmargin}{\strut #1\strut}}
\begin{algorithmic}[1]
\State Initialization: $\mathbf{c}^{temp}_1 = \mathbf{c}^{temp}_2 = \cdots = \mathbf{c}^{temp}_N = \mathbf{0}$, $F = 0$.
\For{$m = 1$ to $M$}
\For{$i = 1$ to $N$}
\State Randomly generate $\mathbf{c}^{temp}_i$ in its domain.
\EndFor
\State Encode $\mathbf{c}^{temp}_1$, $\cdots$, $\mathbf{c}^{temp}_N$ as chromosome $\mathbf{CH}^{temp}_m$.
\EndFor
\State $\eth_0 \gets \{\mathbf{CH}^{temp}_1,\mathbf{CH}^{temp}_2,\cdots,\mathbf{CH}^{temp}_M\}$.
\While{the termination condition is not met}
\For{$m = 1$ to $M$}
\If{$\mathrm{Fit}(\mathbf{CH}^{temp}_m) > F$}
\State $\mathbf{CH}^{best} \gets \mathbf{CH}^{temp}_m$.
\State $F \gets \mathrm{Fit}(\mathbf{CH}^{temp}_m)$.
\EndIf
\EndFor
\For{$m = 1$ to $M$}
\parState{%
Perform recombination operation on $\mathbf{CH}^{temp}_m$ and $\mathbf{CH}^{best}$ to get $\mathbf{CH}^{new}_m$.}
\State Perform mutation operation on $\mathbf{CH}^{new}_m$.
\State $\mathbf{CH}^{temp}_m \gets \mathbf{CH}^{new}_m$
\EndFor
\EndWhile
\State $F \gets 0$.
\State $\eth \gets \{\mathbf{CH}^{temp}_1,\mathbf{CH}^{temp}_2,\cdots,\mathbf{CH}^{temp}_M\}$.
\For{$m = 1$ to $M$}
\If{$\mathrm{Fit}(\mathbf{CH}^{temp}_m) > F$}
\State $\mathbf{CH}^{best} \gets \mathbf{CH}^{temp}_m$.
\State $F \gets \mathrm{Fit}(\mathbf{CH}^{temp}_m)$.
\EndIf
\EndFor
\State Decode $\mathbf{CH}^{best}$ to get $\mathbf{c}_1$, $\mathbf{c}_2$, $\cdots$, $\mathbf{c}_N$.
\State \Return $\mathbf{c}_1$, $\mathbf{c}_2$, $\cdots$, $\mathbf{c}_N$
\end{algorithmic}
\end{algorithm}
\end{figure}

\subsection{Cameras Deployment Using Improved Genetic Algorithm}
The standard genetic algorithm is widely applied to optimize the discrete cost function due to its global search characteristic and strong robustness. However, a drawback of the standard genetic algorithm is the low convergence speed due to the monotonousness of the cost function in iteration process cannot be guaranteed. In order to deal with this disadvantage, an improved genetic algorithm is proposed to guarantee the monotonousness of the cost function in iterations and hence speed up the iteration process.

In the improved genetic algorithm, each feasible solution is encoded as a string called a chromosome, and several chromosomes form a population, as same as that in the standard genetic algorithm. The parameters of the improved genetic algorithm are shown in Table. \ref{t2}. The optimization details for coding and decoding, fitness function, selection operation, recombination operation and mutation operation are introduced as follows.

\begin{figure}[!t]
\centering
\includegraphics[scale=1]{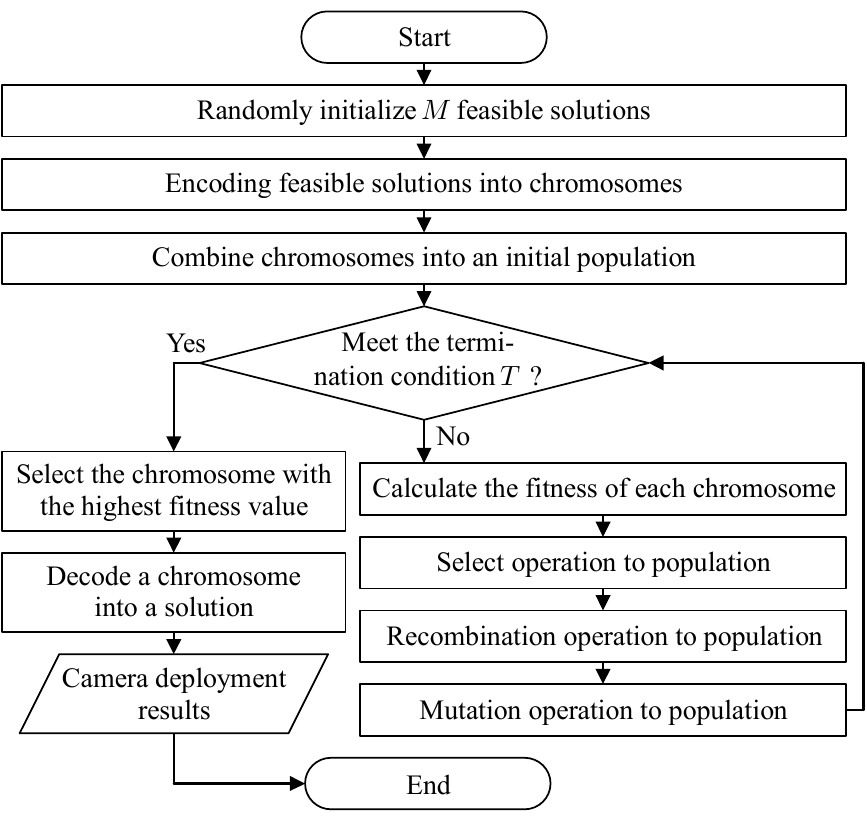}
\caption{Flowchart of improved genetic algorithm.}
\label{S5_IGA_process}
\end{figure}

\subsubsection{Coding and Decoding}
Coding refers to converting a feasible solution into a chromosome, and decoding is an inverse conversion of coding. Each gene in the improved genetic algorithm is represented by a floating-point number. Suppose $\{\mathbf{c}_1,\mathbf{c}_2,\cdots,\mathbf{c}_N\}$ is a feasible solution, then the encoded chromosome is
\begin{equation}
\label{CH}
\begin{aligned}
\mathbf{CH} = \{&x_1,y_1,z_1,\alpha_1,\beta_1,\gamma_1,\\&x_2,y_2,z_2,\alpha_2,\beta_2,\gamma_2,\cdots \\
&\cdots,x_N,y_N,z_N,\alpha_N,\beta_N,\gamma_N\}
\end{aligned}
\end{equation}

Let $\eth = \{\mathbf{CH}_1,\mathbf{CH}_2,\cdots,\mathbf{CH}_M\}$ be a population, where $\mathbf{CH}_m$ is the $m^{th}$ chromosome ($m=1,2,\cdots, M$), and the $t^{th}$ population is represented by $\eth_t$.

\subsubsection{Fitness Function}
The fitness function is used to calculate the fitness of a chromosome which represent the quality of a chromosome. In this paper, the fitness function is defined as same as the cost function, namely,
\begin{equation}
\label{F}
\mathrm{Fit}(\mathbf{CH}) = \mathcal{H}(\mathbf{c}_1,\cdots,\mathbf{c}_N,\mathbf{p}_1,\cdots,\mathbf{p}_K)
\end{equation}

\subsubsection{Selection Operation}
The chromosome in the population with highest fitness is selected as the survival chromosome, and the remaining chromosomes constitute the recombinant pool.

\subsubsection{Recombination Operation}
Randomly generate a positive integer $\Upsilon$ satisfying $\Upsilon\in[\Upsilon_{min},\Upsilon_{max}]$. When two chromosomes are to be recombined, a continuously gene fragment of length $\Upsilon$ is randomly selected from the chromosome with high fitness, and the gene fragment of another chromosome at the corresponding position is replaced to form a new chromosome.

\subsubsection{Mutation Operation}
Mutation refers to a stochastic change of a gene, every gene on a chromosome has a certain probability of mutation. If a gene on a chromosome has not been replaced in recombination operation, the probability of mutation is $\Psi$, otherwise the probability of mutation is zero.

\begin{figure}[!t]
\centering
\includegraphics[scale=1]{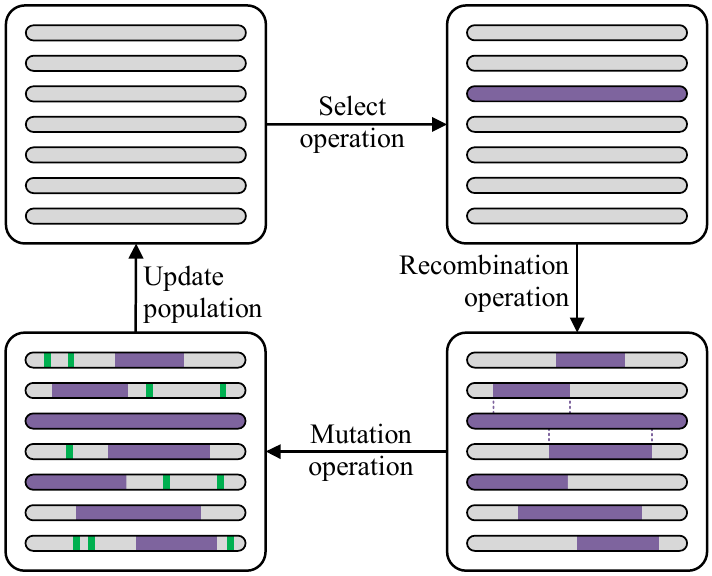}
\caption{Population update operations.}
\label{S5_population_update}
\end{figure}

The flowchart and the detailed procedures are described in Fig. \ref{S5_IGA_process} and Algorithm \ref{algorithm1}, respectively. The update operations of population is shown in Fig. \ref{S5_population_update}, in which the gray bar represents the chromosome, the purple bar is the survival chromosome, and the green bar is the mutated gene. The parameters of the improved genetic algorithm are included in an eight-tuple:
\begin{equation}
\label{IGA}
IGA = (W,\mathrm{Fit},\eth_0,M,\Upsilon_{min},\Upsilon_{max},\Psi,T)
\end{equation}
where $W$ is the coding scheme of the chromosome and $\eth_0$ is the initial population. Since the chromosome with the highest fitness in the current population is remained in next population, the non-decrement of the overall fitness during the iteration process is guaranteed, and hence the convergence speed of the algorithm is greatly accelerated.

\section{Simulation and Experiment}
\label{section6}
Some simulations and experiments are shown in this section to verify the contents in the above sections, which are summarized in three aspects, (a) the quality of the fused image is positive correlated with the fused coverage strength, (b) the improved genetic algorithm is valid for the constructed cost function and has good performance and fast convergence rate, and (c) the proposed radial coverage strength and multi-camera network deployment approach can be applied to the real case. In addition, two simplified fusion coverage strength algorithms are also proposed to speed up the calculation of the cost function.

\begin{figure}[!t]
\centering
\includegraphics[scale=1]{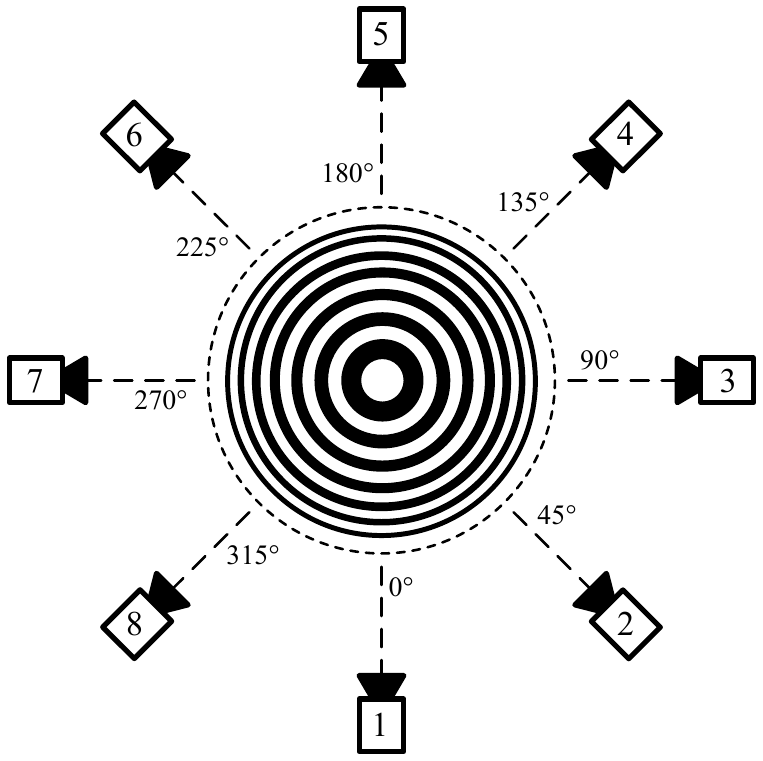}
\caption{Omnidirectional test image and camera placement.}
\label{S6_image_test}
\end{figure}

\begin{figure}[!t]
\centering
\includegraphics[scale=1]{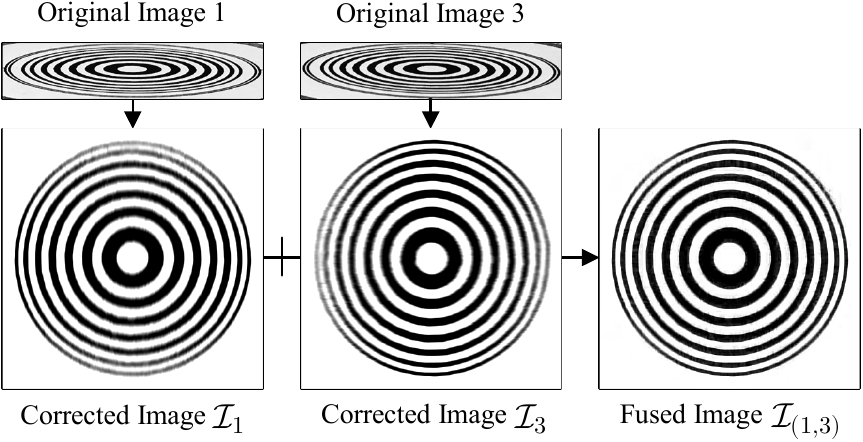}
\caption{An example of image correction and image fusion.}
\label{S6_image_test_example}
\end{figure}

\subsection{Simplified Fused Coverage Strength Algorithm}
Generally speaking, in the aforementioned fused coverage strength algorithm, $N^2$ times calculations are needed to obtain the fused coverage strength. The fact is that the elapsed time of completing the fused coverage strength calculations of all the triangle pieces is long for the large sized camera network. Since decreasing the number of the triangle pieces will reduce the accuracy of the object model, a simplified fused coverage strength algorithm are proposed to speed up calculations.

For each triangle piece, a principal camera and an auxiliary camera are chosen  according to the following rules. For $\tau_k$, let $c_{\hat{i\,}\!}$ be the principal camera and $c_{\hat{j\,}\!}$ be the auxiliary camera, where $\hat{i\,}\!,\hat{j\,}\!\in \{1,2,\cdots, N\}$. The principal camera can be selected according to the following two methods:
\begin{equation}
\label{hat_i_1}
\hat{i\,}\! = \mathop{\arg\max}_{i}\,\|\mathbf{Cs}^{ci}_k\|
\end{equation}
called the coverage-strength-based method (CSBM) or
\begin{equation}
\label{hat_i_2}
\begin{aligned}
\hat{i\,}\! = &\mathop{\arg\max}_{i}\,\mathcal{H}(\mathbf{c}_i,\mathbf{c}_i,\mathbf{p}_1,\cdots,\mathbf{p}_K)\\
&~~~~~~s.t. ~~\mathbf{c}_i\in\Gamma,\mathbf{p}_k\in\Omega
\end{aligned}
\end{equation}
called the recognized-area-based method (RABM). The auxiliary camera is chosen through following methods:
\begin{equation}
\label{hat_j}
\hat{j\,}\! = \mathop{\arg\max}_{j}\,Cs_k(c_{\hat{i\,}\,\!},c_j)
\end{equation}
Then the fused coverage strength of $\mathbf{p}_k$ under camera network can be formulated as
\begin{equation}
\label{Cs_p_k_2}
Cs(\mathbf{p}_k) = Cs_k(c_{\hat{i\,}\,\!},c_{\hat{j\,}\!})
\end{equation}

The simplified fused coverage strength algorithm with coverage-strength-based method and recognized-area-based method will lead to different results of the fusion coverage strength. Both of them can greatly reduce the number of calculations, which is an approximation of the fusion coverage strength algorithm proposed in Section \ref{section4}. In addition, it is pointed out that if $c_{\hat{i\,}\!}$ is not unique, it is necessary to calculate the fused coverage strength with different principal cameras respectively, and select the largest one as the final fused coverage strength.

\subsection{Experimental Results of Image Fusion}

In this experiment, an omnidirectional black-and-white circles image is used for detection of the quality of the image due to its abundant edges. This tested image is printed out and placed on a platform. We use Cannon 50-D camera to shoot the tested image from $8$ different directions, and each angle between the adjacent shooting directions are set to be $45$ degrees. To ensure the rigor of the experiment, the depth distances and the view angles between the camera and the tested image in $8$ directions are strictly equivalent. The imaging is complete and clear in the camera from all $8$ directions. This scene is shown in Fig. \ref{S6_image_test}.

\begin{table}[!t]
\renewcommand{\arraystretch}{1.3}
\caption{Experimental Results of Image Fusion}
\label{t3}
\centering
\begin{tabular}{lcccc}
\hline\hline
\specialrule{0em}{0pt}{2pt}
Fused Image & $\mathcal{I}_{(1,1)}$ & $\mathcal{I}_{(1,2)}$ & $\mathcal{I}_{(1,3)}$ & $\mathcal{I}_{(1,4)}$\\
\specialrule{0em}{2pt}{0pt}
\hline
\specialrule{0em}{2pt}{0pt}
Coverage Strength & 2.000 & 2.449 & 2.828 & 2.449 \\
RMSE & 51.266 & 48.903 & 46.049 & 49.090 \\
PSNR & 13.934 & 14.344 & 14.866 & 14.311 \\
\specialrule{0em}{2pt}{0pt}
\hline
\specialrule{0em}{0pt}{2pt}
Fused Image & $\mathcal{I}_{(1,5)}$ & $\mathcal{I}_{(1,6)}$ & $\mathcal{I}_{(1,7)}$ & $\mathcal{I}_{(1,8)}$ \\
\specialrule{0em}{2pt}{0pt}
\hline
\specialrule{0em}{2pt}{0pt}
Coverage Strength & 2.000 & 2.449 & 2.828 & 2.449 \\
RMSE & 51.172 & 49.229 & 46.070 & 48.747 \\
PSNR & 13.950 & 14.286 & 14.862 & 14.372 \\
\specialrule{0em}{2pt}{0pt}
\hline\hline
\end{tabular}
\end{table}

\begin{table}[!t]
\renewcommand{\arraystretch}{1.3}
\caption{Recognition Ratio With Heuristic Method}
\label{t4}
\centering
\begin{tabular}{lcccc}
\hline\hline
\specialrule{0em}{0pt}{2pt}
Cameras & 5 & 6 & 7 & 8 \\
\specialrule{0em}{2pt}{0pt}
\hline
\specialrule{0em}{2pt}{0pt}
Recognized Ratio & 83.70\% & 89.37\% & 92.13\% & 95.21\% \\
\specialrule{0em}{2pt}{0pt}
\hline\hline
\end{tabular}
\end{table}

\begin{figure*}[!t]
\centering
\subfloat[Improved genetic algorithm with coverage-strength-based method]{\includegraphics[scale=1]{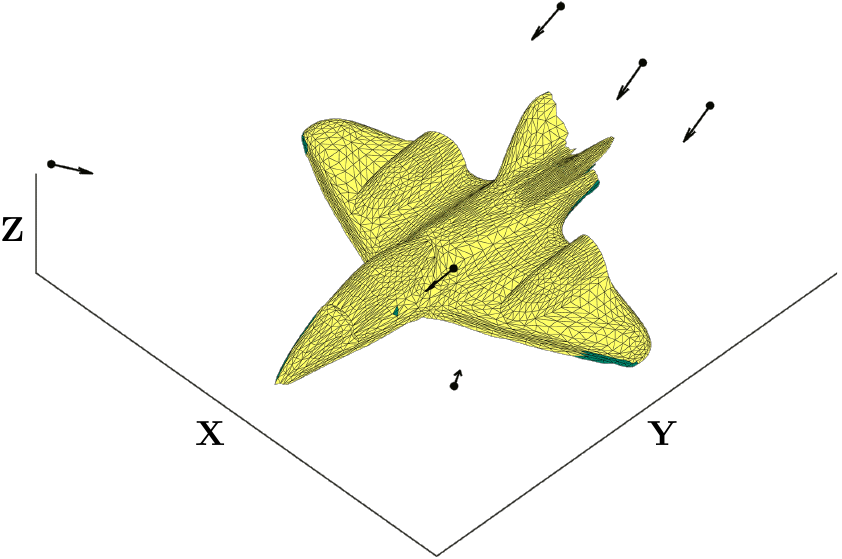}
\label{S6_recognized_results_a}}
\hfil
\subfloat[Heuristic algorithm with coverage-strength-based method]{\includegraphics[scale=1]{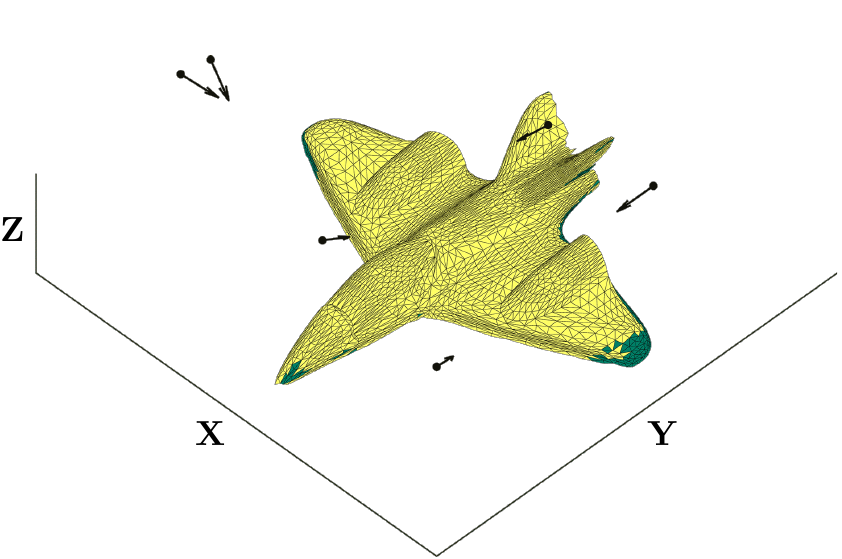}
\label{S6_recognized_results_b}}
\hfil
\subfloat[Improved genetic algorithm with recognized-area-based method]{\includegraphics[scale=1]{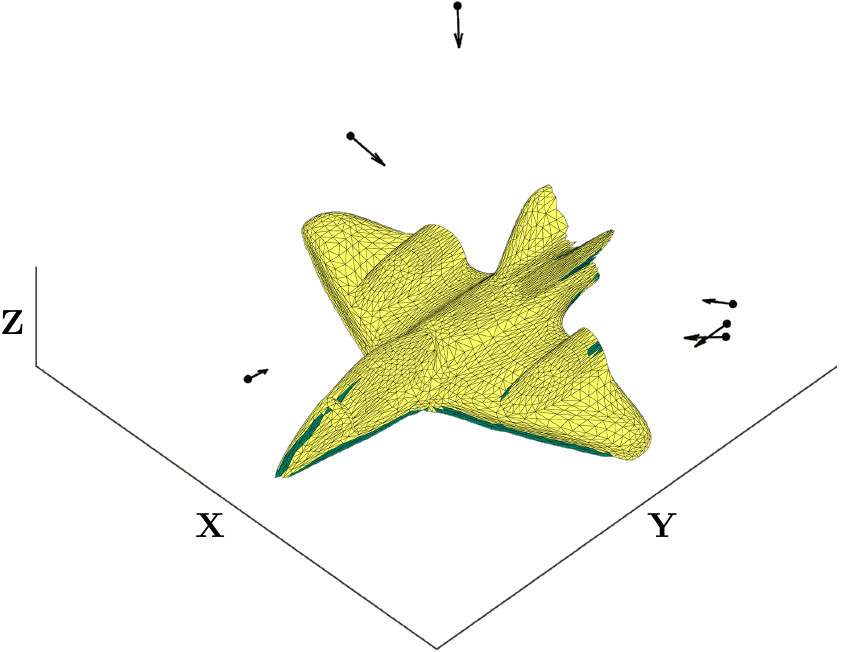}
\label{S6_recognized_results_c}}
\hfil
\subfloat[Standard genetic algorithm with coverage-strength-based method]{\includegraphics[scale=1]{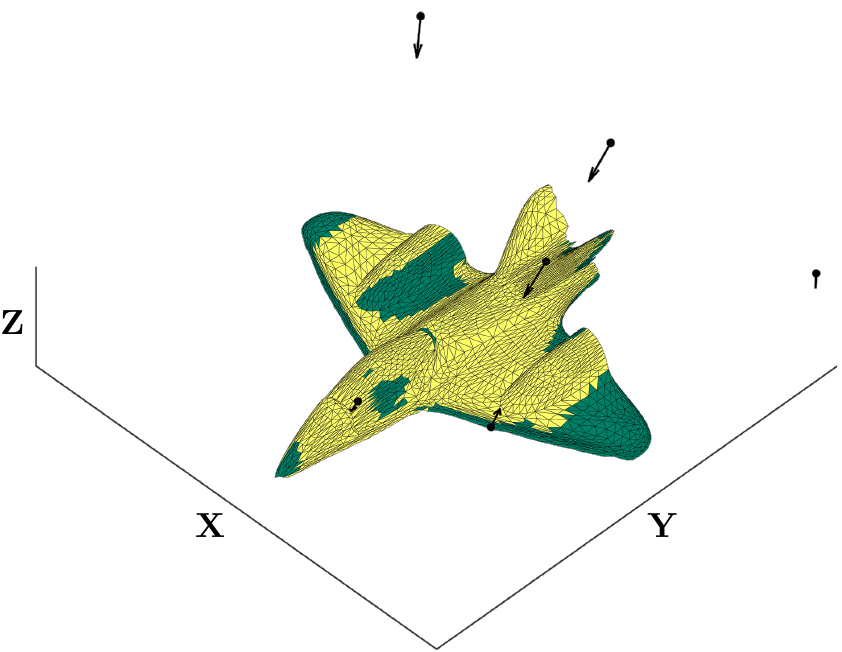}
\label{S6_recognized_results_d}}
\caption{The coverage results with different algorithms.}
\label{S6_recognized_results}
\end{figure*}

The registration processes are operated on the captured $8$ images to obtain the corrected images which are marked as $\mathcal{I}_1,\mathcal{I}_2,\cdots,\mathcal{I}_8$. The original image is processed into a reference image $\mathcal{I}_R$ of the same size as the corrected images. $\mathcal{I}_1$ is selected as a basic image and make image fusion operation on it with all the $8$ corrected images, then $8$ fused images ($\mathcal{I}_{(1,1)},\mathcal{I}_{(1,2)},\cdots,\mathcal{I}_{(1,8)}$) are obtained. It is noted that many image fusion techniques have been tested in literatures, such as high pass filtering \cite{aiazzi2002context}, discrete wavelet transform \cite{yocky1995image, ranchin2003image}, uniform rational filter bank \cite{blanc1998using} and Laplacian pyramid \cite{wang2011multi}. Wavelet transformation method is used in this experiment for image fusion. Each fused image is compared with $\mathcal{I}_R$ to calculate the root mean square error (RMSE) and the peak signal to noise ratio (PSNR). The lower the RMSE or the higher the PSNR, the better the quality of the image. Fig. \ref{S6_image_test_example} shows an example of the procedure of the image registration and fusion.

The qualities of $8$ fused images together with the corresponding fused coverage strength are stated in Table \ref{t3}. It can be seen that the quality of the fused image is positive correlated with the fused coverage strength. Due to the value of the coverage strength of the camera in each direction are equivalent, the fused coverage strength totally depend on the orientations between cameras. It is straightforward that the fused coverage strength is largest when the angle between two shooting directions is $90$ degree, while the fused image quality is also the best.

\begin{table}[!t]
\renewcommand{\arraystretch}{1.3}
\caption{Evaluation Metrics for Different Algorithms and Methods}
\label{t5}
\centering
\begin{tabular}{lcccc}
\hline\hline
\specialrule{0em}{0pt}{2pt}
Algorithms / Methods & \makecell{Recognized \\ratio} & \makecell{Average \\coverage strength} & \makecell{Average \\resolution} \\
\specialrule{0em}{2pt}{0pt}
\hline
\specialrule{0em}{0pt}{2pt}
IGA / CSBM & 94.92\,\% & 1.744 & 1.507 \\
IGA / RABM & 92.34\,\% & 1.729 & 1.651 \\
HA / CSBM & 89.37\,\% & 1.704 & 1.482 \\
MGA / CSBM & 83.66\,\% & 1.737 & 1.485 \\
SGA / CSBM & 69.36\,\% & 1.306 & 1.143 \\
\specialrule{0em}{2pt}{0pt}
\hline\hline
\end{tabular}
\end{table}

\subsection{Simulation Results of Cameras Deployment with IGA}

In this subsection, the effectiveness of the proposed radial coverage strength and improved genetic algorithm for multi-camera deployment is demonstrated by a simulation example. The robustness and the fast convergence of the improved genetic algorithm are also shown. In addition, different optimization methods and fusion coverage strength methods are compared.

In the simulation, an aircraft shell model is employed as the object model, and the camera network are deployed to cover the top and the flank of the aircraft shell model. To simplify the calculations, The heights of all the cameras are limited at $z_i = 1.6$ m, such that all cameras have only five degrees of freedom to be changed. The parameters of the aircraft shell model and the cameras are set as follows: The aircraft shell model is partitioned into $K = 4589$ triangle pieces. Each camera has the same intrinsic parameters. The lens focal length is $f = 5$ mm, the horizontal and vertical pixel dimensions are $s_u = s_v = 0.0053$ mm/pixel, and the principle point is
$
\mathbf{o} = [\begin{IEEEeqnarraybox*}[][c]{,c/c,}
800 & 600
\end{IEEEeqnarraybox*}]^\mathrm{T}
$
in pixel, with the image width being $w = 1600$ and the image height being $h = 1200$ in pixel. The effective aperture diameter of the optical lens is $d_a = 5$ mm, and the focusing distance is $d_s = 1200$ mm. The diameter of permissible circle of confusion is $\delta = 5$ pixel, with the coverage strength threshold being $thold = 1$. On the basis of these parameters, the other parameters such as the view angles $\varphi_l$, $\varphi_l$, $\varphi_t$ and $\varphi_b$, the near depth of field $d_n$ and the far depth of field $d_f$ can be calculated directly.

The heuristic algorithm (HA) are applied first to determine the number of cameras, that is, deploy the cameras one by one, and each camera is deployed to its best pose at current. The details of this method can be found in \cite{zhang20153}. The relationship of the recognized ratio and the number of the camera network is illustrated in Table \ref{t4}, then $6$ cameras are selected for further optimization.

\begin{figure}[!t]
\centering
\includegraphics[scale=1]{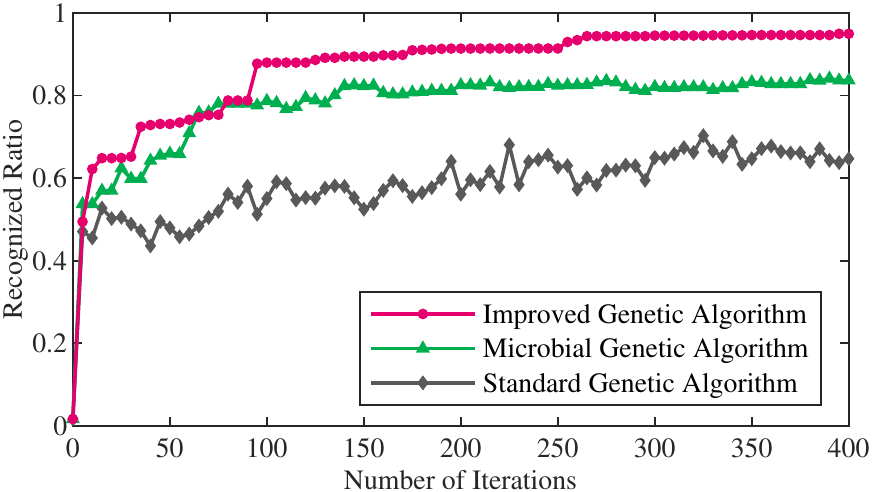}
\caption{Comparison of the recognized ratios with different methods.}
\label{S6_plot_results}
\end{figure}

The parameters of the improved genetic algorithm are set as follows: The size of the population is $M=20$, the length of the chromosome is $L=5\times N=30$ due to each camera has five degrees of freedom. The minimum and maximum length of recombination are $\Upsilon_{min}=11,\Upsilon_{max}=19$, respectively, and the mutation probability of each gene is $\Psi=0.2$. Moreover, to be fair for comparison of different algorithms, all the optimization algorithms in this subsection are set to terminate at the $400^{th}$ iteration.

The following aspects are considered as the performances of the multi-cameras network deployment:
\begin{enumerate}[\IEEEsetlabelwidth{12)}]
\item
\emph{Recognized Ratio:} The proportion of the recognized triangle pieces to all the triangle pieces in the 3-D object model.
\item
\emph{Average Coverage Strength:} The average of the fused coverage strength for all triangle pieces.
\item
\emph{Average Resolution:} The average of the value of resolution for all triangle pieces in pixel/mm.
\end{enumerate}

To speed up the calculation, the simplified coverage strength algorithm is applied in simulations. The comparison of different algorithms and methods for the performances optimization are shown in Fig. \ref{S6_recognized_results} and Table \ref{t5}. The results of using standard genetic algorithm and heuristic algorithm are also compared. In Fig. \ref{S6_recognized_results}, the black dots and arrows indicate the positions and the directions of the cameras, the recognized triangle pieces are marked as yellow while the unrecognized ones are marked as green. It can be seen that the performances of the multi-camera deployment with the improved genetic algorithm are better than those with standard genetic algorithm and heuristic algorithm. In the case of using the same optimization algorithm, the recognized ratio of coverage-strength-based method is higher than recognized-area-based method. The coverage performances are shown in Table \ref{t5}, and the relationship between the recognized ratio and iteration are demonstrated in Fig. \ref{S6_plot_results}, with different optimization algorithms and all using recognized-area-based method. As can be seen, the recognized ratio of the proposed improved genetic algorithm converges fastest among them.

\begin{figure}[!t]
\centering
\includegraphics[scale=1]{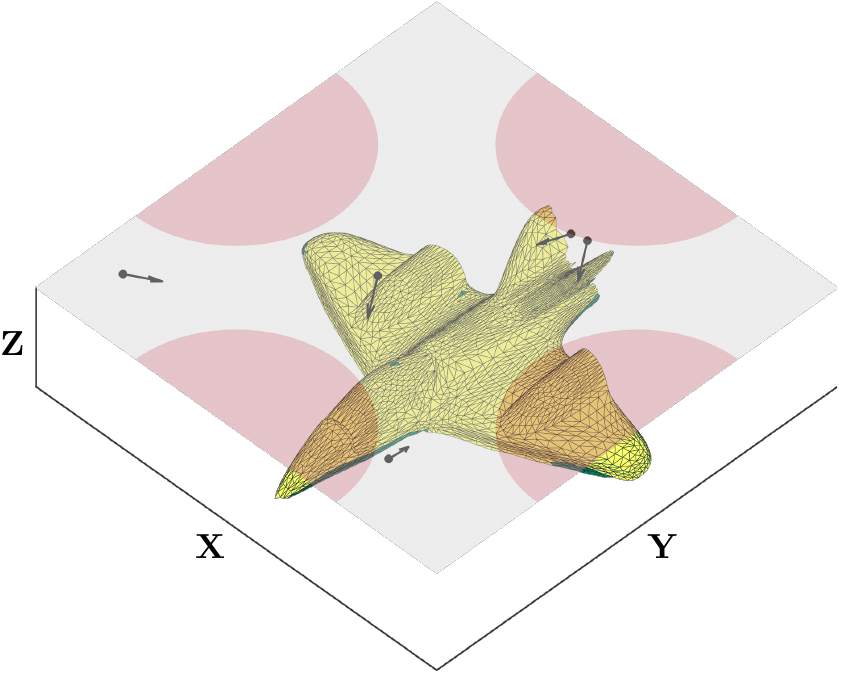}
\caption{Improved genetic algorithm with constraints.}
\label{S6_recognized_results_constraints}
\end{figure}

In addition, a case with some external obstacles existing is also taken into account in this subsection. Some external obstacles existing around the object model are assumed such that the locations of some cameras in Fig. \ref{S6_recognized_results} are not available for placement. Under this constraint, the final deployment of the cameras is shown in Fig. \ref{S6_recognized_results_constraints} using improved genetic algorithm with coverage-strength-based method, in which the pink semicircles indicate the external obstacles. It can be seen that the performance of coverage is still perfect although the external obstacles exist to obstruct the deployment.

\begin{figure*}[!t]
    \centering
    \begin{minipage}[t]{0.497\textwidth}
        \centering
        \includegraphics[width=8.8cm]{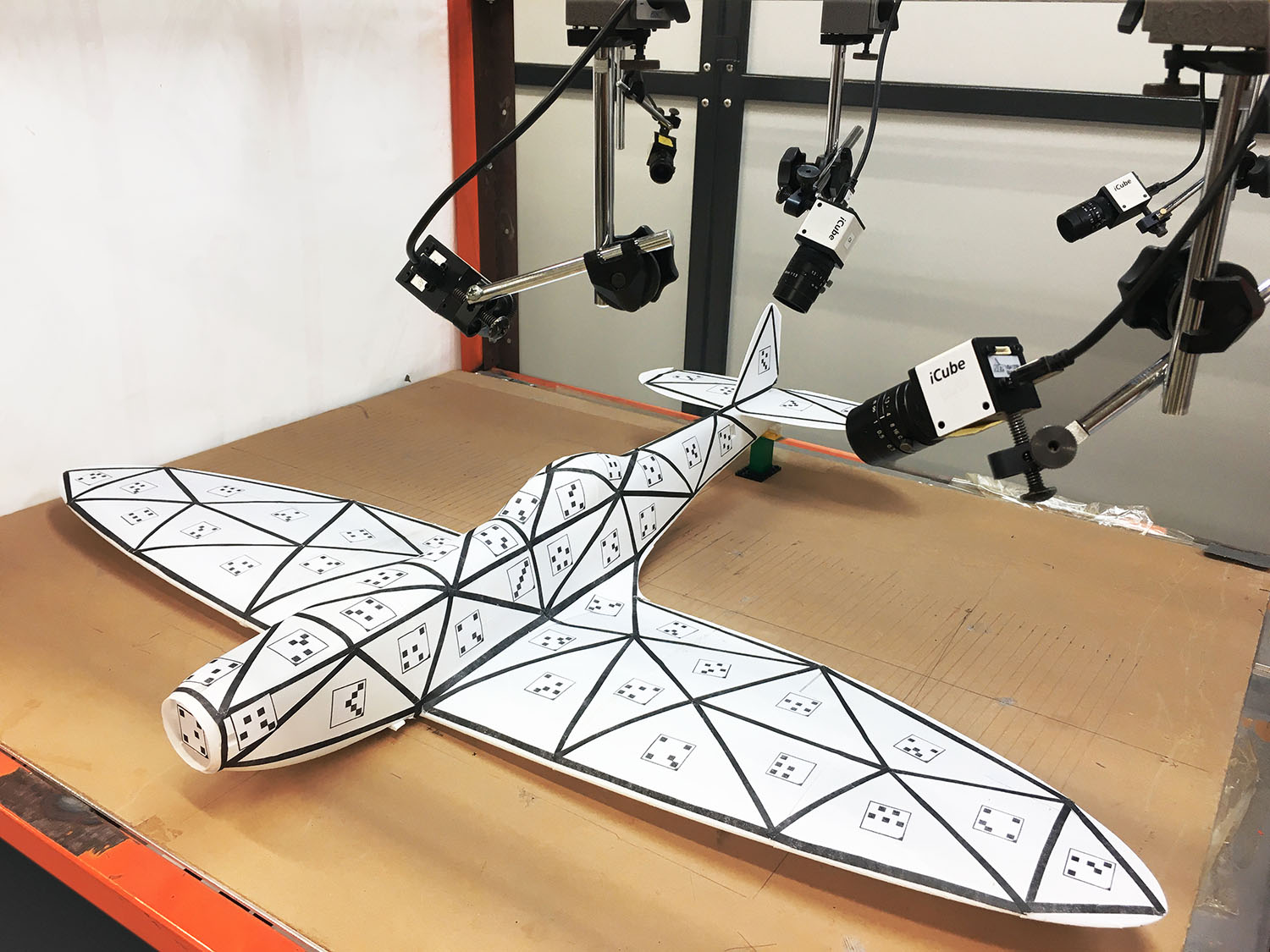}
        \caption{Experiment platform}
        \label{S6_experiment_picture}
    \end{minipage}
    \begin{minipage}[t]{0.497\textwidth}
        \centering
        \includegraphics[scale=1]{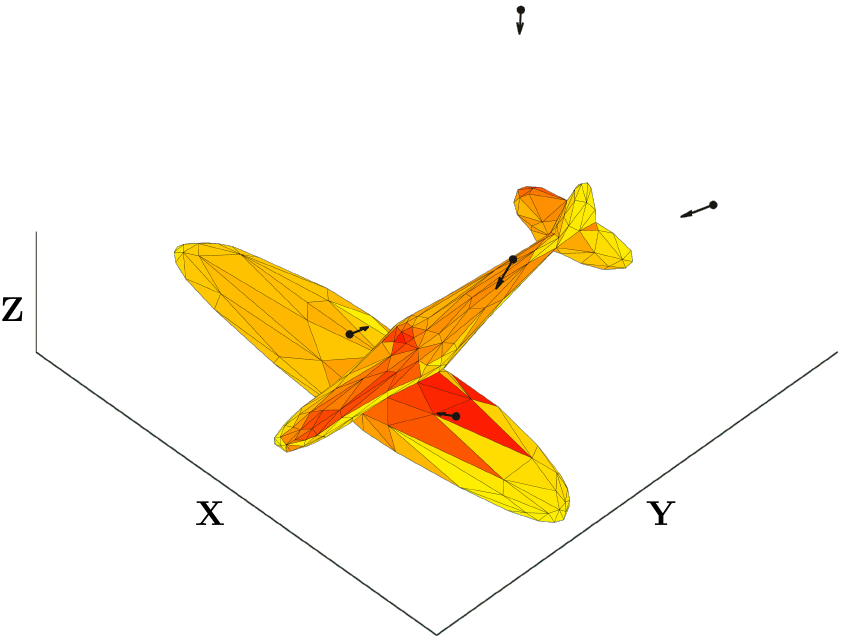}
        \caption{The coverage results in simulation using IGA with CSBM}
        \label{S6_experiment_deployment}
    \end{minipage}
\end{figure*}

%
%

\begin{figure*}[!t]
\centering
\includegraphics[scale=1]{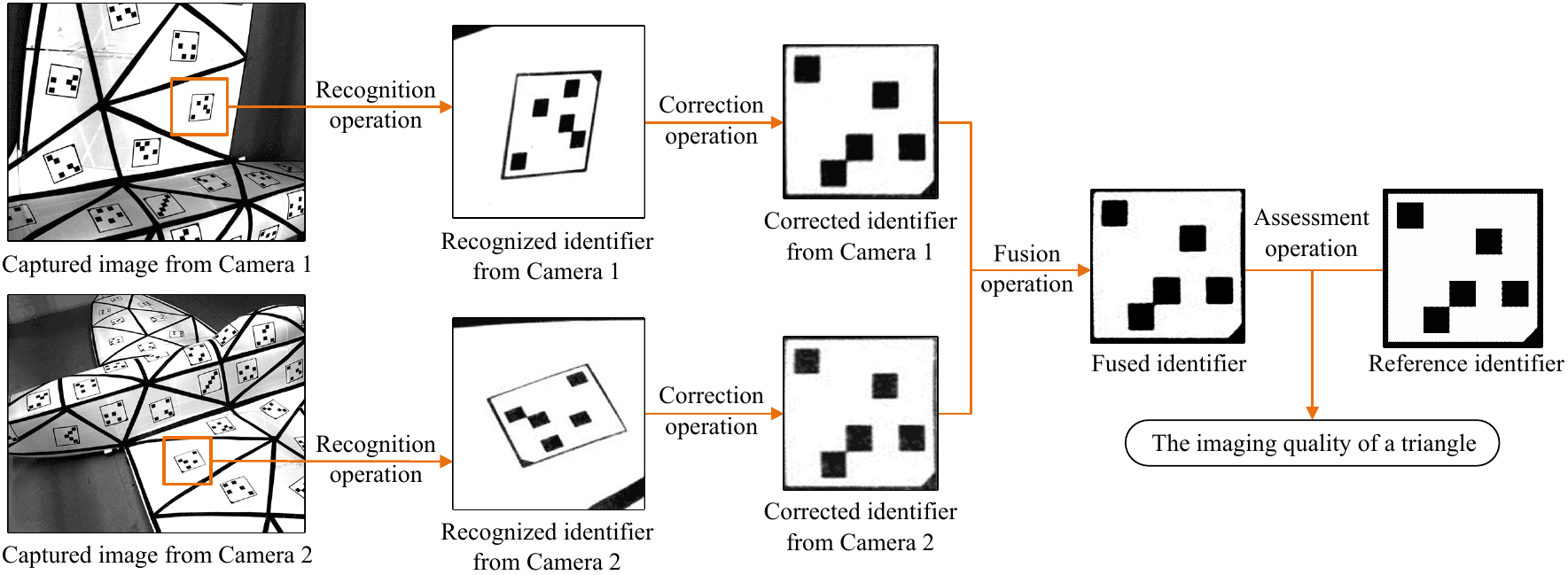}
\caption{An example of imaging quality assessment for an identifier.}
\label{S6_experiment_example}
\end{figure*}

\begin{figure*}[!t]
\centering
\subfloat[Simulation result]{\includegraphics[scale=0.25]{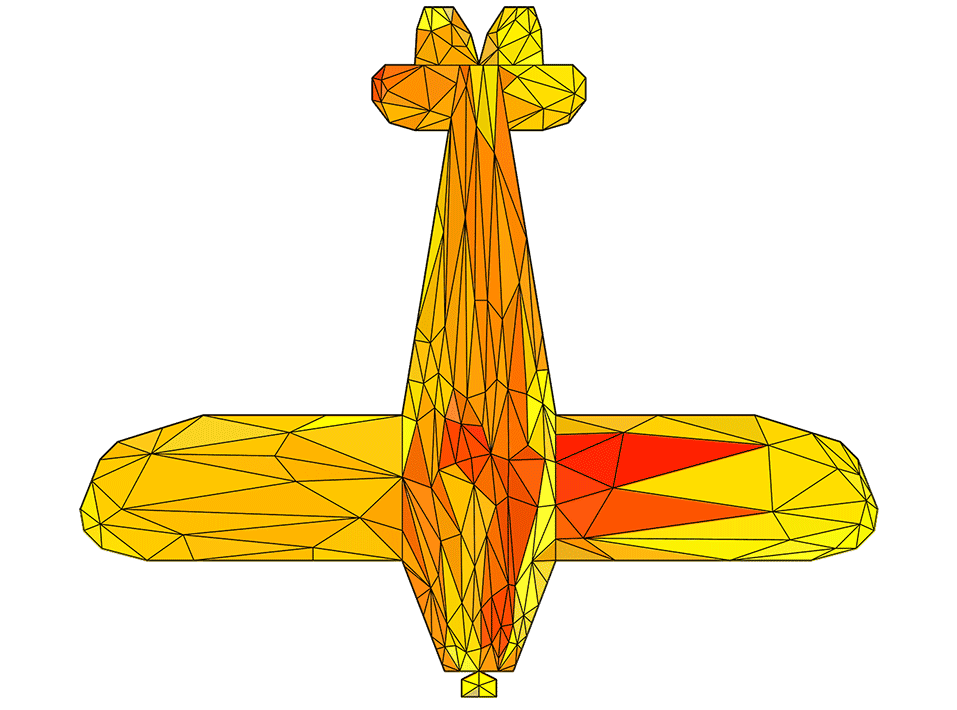}
\label{S6_experiment_compare_a}}
\hfil
\subfloat[Experiment result]{\includegraphics[scale=0.25]{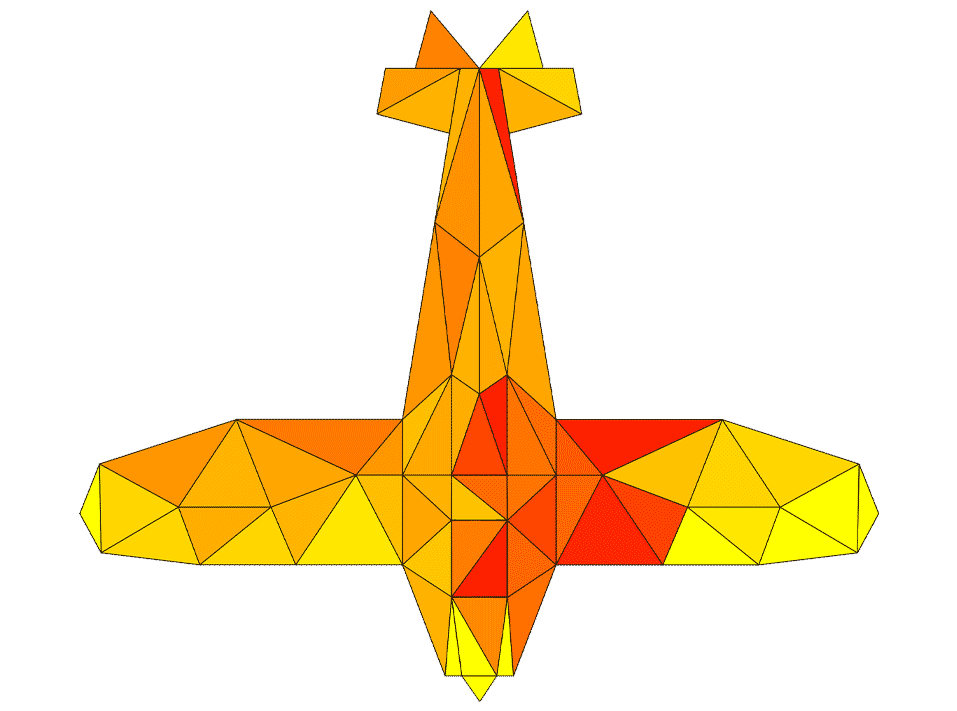}
\label{S6_experiment_compare_b}}
\caption{Comparison of coverage performances in simulation and experiment for UV unwrapped mesh images}
\label{S6_experiment_compare}
\end{figure*}

\subsection{Experimental Results of Cameras Deployment}
A real experiment is designed and operated on a platform with five cameras to further verify the effectiveness of the proposed approach. The objective to be covered is a small sized real aircraft shell model manufactured by 3-D print. Each camera on the platform has the same intrinsic parameters. The lens focal length is $f = 15.2$ mm, the horizontal and vertical pixel dimensions are $s_u = s_v = 0.0053$ mm/pixel, the principle point is
$
\mathbf{o} = [\begin{IEEEeqnarraybox*}[][c]{,c/c,}
640 & 512
\end{IEEEeqnarraybox*}]^\mathrm{T}
$
in pixel, with the image width being $w = 1280$ and the image height being $h = 1024$ in pixel. The effective aperture diameter of the optical lens is $d_a = 3.8$ mm, and the focusing distance is $d_s = 1000$ mm. The heights of the five cameras are limited at 30 cm. Some pre-processing are accompanied on the aircraft model. The aircraft model is partitioned in to $66$ triangle pieces, and a series of identifiers are pasted on each triangle piece. The identifier is actual an image consisting of same sized black squares with same number of graphic edges. Each identifier is different for the sake of distinguishing each triangle piece on the objective. Another important purpose of using these identifiers is to analyze the quality of the images captured by cameras. The experimental platform and the pre-processed aircraft model are shown in Fig. \ref{S6_experiment_picture}.

The placements of cameras including the positions and orientations are calculated by simulation, and the values of the coverage strength are obtained correspondingly. The triangle pieces with higher coverage strength values are marked by darker color. The simulation deployment result is shown in Fig. \ref{S6_experiment_deployment}, then the cameras are placed on the platform based on the simulation result, and images will be captured by the cameras. Next, the images are processed to assess the qualities. Specifically, the identifiers on the images are identified and corrected at beginning. If an identifier is captured by multiple cameras, the images involving it will be fused after correction. The corrected and fused images will be compared with the reference images to calculate RMSE to assess the qualities.  An example of imaging quality assessment for an identifier is shown in Fig. \ref{S6_experiment_example}.


To reveal whether the calculated fused coverage strength can represent the qualities of the real images captured by the cameras, the simulation result is compared with the real experiment result. Considering the errors caused by different partitions of the triangle pieces in simulation and real experiment, UV unwrapping is performed on the 3-D model such that each triangle piece of the 3-D model can be seen from a 2-D mesh image. The results of the simulation and experiment are shown in Fig. \ref{S6_experiment_compare_a} and \ref{S6_experiment_compare_b}, respectively. In Fig. \ref{S6_experiment_compare}, the areas with lower coverage strength or lower image qualities are marked as lighter yellow, while the areas with higher coverage strength or higher image qualities are marked as darker red. From the comparison, the trend of the color change on the triangle pieces are similar for simulation and experiment, which validates the effectiveness of the proposed fused coverage approach for cameras network deployment.

\section{Conclusion}
\label{section7}
A multi-camera network deployment approach is proposed in this paper to solve the coverage problem for 3-D object model. A new criterion called radial coverage vector is proposed to characterize the visual sensing performance, which takes the resolution, FOV, focus and occlusion into account, and also the relative orientation between the camera and triangle piece. Then the effective radial coverage vector and fusion vector are decomposed from the radial coverage vector to calculate the fused coverage strength through the proposed fused coverage strength algorithm. An improved genetic optimization algorithm is applied to solve the final multi-camera deployment, while two simplified method are presented to speed up calculations. In the end, some simulations and experiments are presented to illustrate the superior performance of the proposed approach.


%



\ifCLASSOPTIONcaptionsoff
  \newpage
\fi



%

\bibliographystyle{IEEEtran}
\bibliography{IEEEabrv,IEEEexample}

%

\vspace*{-2\baselineskip}

\begin{IEEEbiography}[{\includegraphics[width=1in,height=1.25in,clip,keepaspectratio]{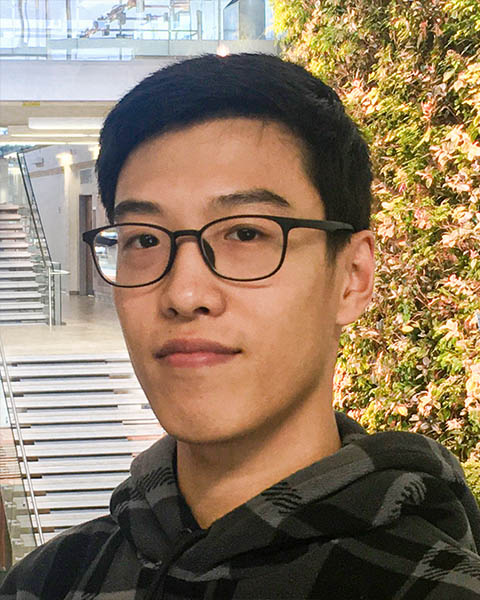}}]{Zike Lei}
received B.Eng. degree in electronic information engineering from Wuhan University of Science and Technology, Hubei, China, in 2017. He is a Ph.D. student in control science and engineering at the School of information Science and Engineering, Wuhan University of Science and Technology. He is currently a visiting scholar at University of Windsor, ON, Canada, since 2019. His research interests include field sensor networks, stereo-camera modeling, and observer-based coverage control.
\end{IEEEbiography}

\vspace*{-2.5\baselineskip}

\begin{IEEEbiography}[{\includegraphics[width=1in,height=1.25in,clip,keepaspectratio]{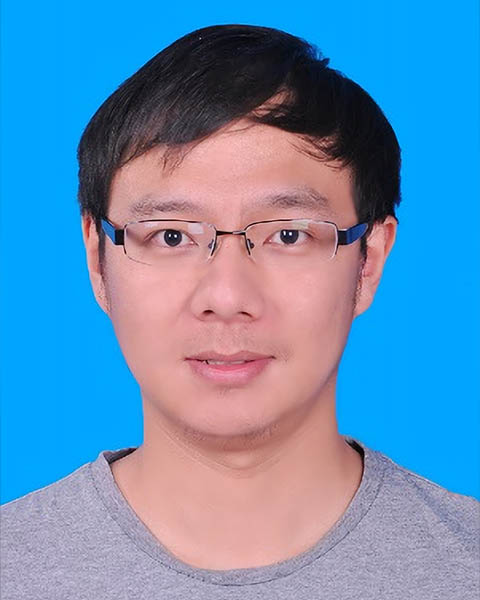}}]{Xi Chen}
received B.S. degree from Huazhong University of Science and Technology, and M.S. degree from Xiamen University, in 2009 and 2012, respectively. He received the Ph.D degree from the University of Newcastle, Australia in 2015. He joined Wuhan University of Science and Technology in 2015, where he is currently an associate professor. His research interests include coverage of sensor network, multi-agent systems and nonlinear system control.
\end{IEEEbiography}

\vspace*{-2.5\baselineskip}

\begin{IEEEbiography}[{\includegraphics[width=1in,height=1.25in,clip,keepaspectratio]{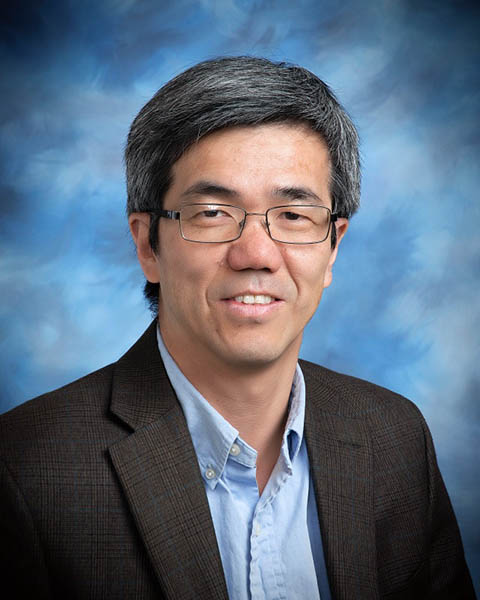}}]{Xiang Chen}
(M’98) received the M.Sc. and Ph.D. degrees from Louisiana State University, Baton Rouge, LA, USA, in 1996 and 1998, respectively, both in systems and control.

Since 2000, he has been with the Department of Electrical and Computer Engineering, University of Windsor, ON, Canada, where he is currently a Professor. His research interests include robust control, vision sensor networks, vision-based control systems, networked control systems, and industrial applications of control theory.
\end{IEEEbiography}

\vspace*{-2.5\baselineskip}

\begin{IEEEbiography}[{\includegraphics[width=1in,height=1.25in,clip,keepaspectratio]{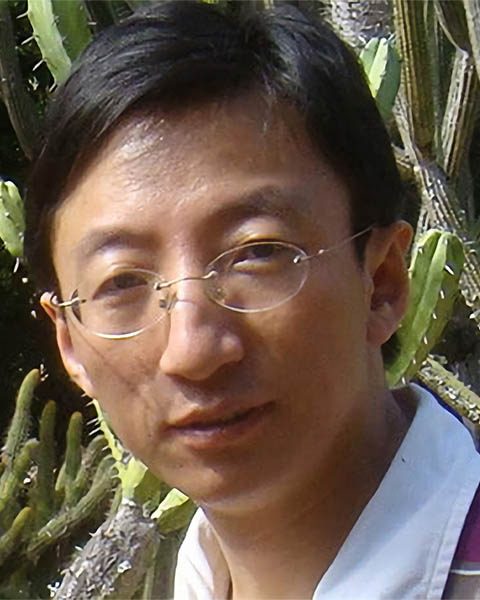}}]{Li Chai}
(S'00-M'03) received the B.S. degree in applied mathematics and the M. S. degree in control science and engineering, both from Zhejiang University, China, in 1994 and 1997 respectively, and the Ph.D. degree in electrical engineering from Hong Kong University of Science and Technology in 2002.

In September 2002, he joined Hangzhou Dianzi University, China. He worked as a postdoctoral research fellow at the Monash University, Australia, from May 2004 to June 2006. In 2008, he joined Wuhan University of Science and Technology, where he is currently a Chutian Chair Professor. He has been a visiting scholar at Newcastle University, Australia, and Harvard University. His research interests include distributed optimization,  filter bank frames, graph signal processing, and networked control systems.

Professor Chai is the recipient of the Distinguished Young Scholar of the National Science Foundation of China. He is currently an associate editor for the Decision and Control.
\end{IEEEbiography}







\end{document}